\documentclass{webofc}
\usepackage[varg]{txfonts}   % Web of Conferences font
\usepackage{slashed} 
\newcommand{\nopieft}{\mbox{$\slashed{\pi}$EFT}~} 

\begin{document}

\title{Onset of $\eta$ nuclear binding}

\author{\firstname{Nir} \lastname{Barnea}\inst{1} \and 
\firstname{Betzalel} \lastname{Bazak}\inst{1} \and 
\firstname{Eliahu} \lastname{Friedman}\inst{1} \and 
\firstname{Avraham} \lastname{Gal}\inst{1}\fnsep\thanks
{Presented by A. Gal (avragal@savion.huji.ac.il) at EXA2017, Vienna, 
Sept. 2017} 
\\ 
\firstname{Ale\v{s}} \lastname{Ciepl\'{y}}\inst{2} \and 
\firstname{Ji\v{r}\'{\i}} \lastname{Mare\v{s}}\inst{2} \and 
\firstname{Martin} \lastname{Sch\"{a}fer}\inst{2} 
}

\institute{Racah Institute of Physics, the Hebrew University, Jerusalem
91904, Israel 
\and 
Nuclear Physics Institute, 250 69 Rez, Czech Republic}

\abstract{ 
Recent studies of $\eta$ nuclear quasibound states by the Jerusalem-Prague 
Collaboration are reviewed, focusing on stochastic variational method self 
consistent calculations of $\eta$ few-nucleon systems. These calculations 
suggest that a minimum value Re$\,a_{\eta N} \approx 1$~fm (0.7~fm) is needed 
to bind $\eta\,^3$He ($\eta\,^4$He).}  

\maketitle

\section{Introduction} 
\label{sec:int} 

\begin{figure}[htb] 
\centering 
\includegraphics[width=0.5\textwidth]{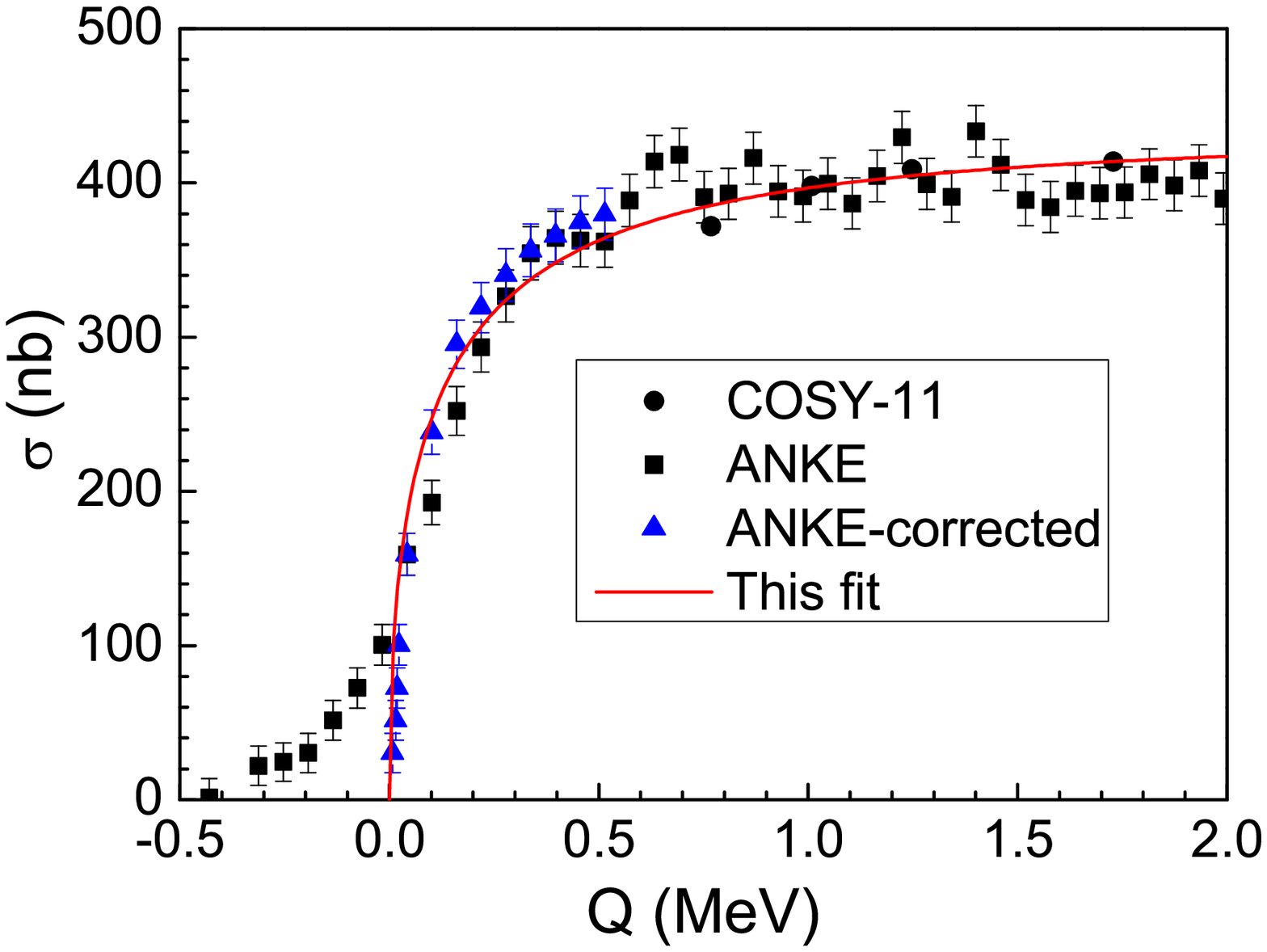}  
\includegraphics[width=0.4\textwidth,height=4.5cm]{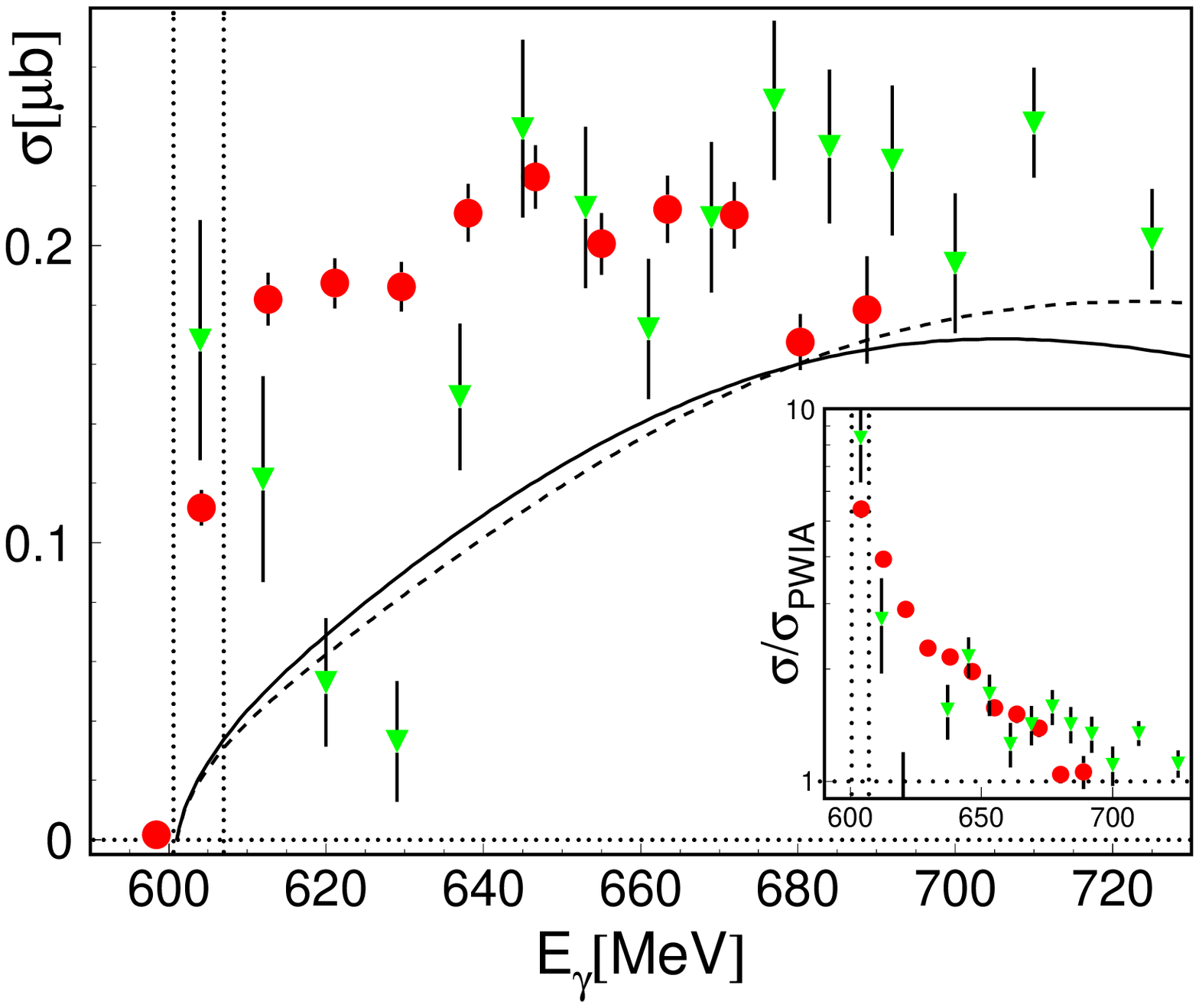}
\caption{Near-threshold $\eta\,^3$He production cross sections. Left: 
$dp\,\to\eta\,^3$He~\cite{Xie17}. Right: $\gamma\,^3$He$\,\to\eta\,^3
$He~\cite{pheron12}.} 
\label{fig:dp} 
\end{figure}

The $\eta N$ near-threshold interaction is attractive, owing to the $N^{\ast}
$(1535) resonance to which the $s$-wave $\eta N$ system is coupled strongly 
\cite{BL85}. This has been confirmed in chiral meson-baryon coupled channel 
models that generate the $N^{\ast}$(1535) dynamically, e.g.~\cite{KSW95}. 
Hence $\eta$ nuclear quasibound states may exist~\cite{HL86} as also suggested 
experimentally by the near-threshold strong energy dependence of the $\eta\,
^3$He production cross sections shown in Fig.~\ref{fig:dp}. However, the $\eta
\,^3$He scattering length deduced in Ref.~\cite{Xie17}, $a_{\eta\,^3{\rm He}}=
[-(2.23\pm 1.29)+i(4.89\pm 0.57)]~{\rm fm}$, although of the right sign of its 
real part, does not satisfy the other necessary condition for a quasibound 
state pole: $-{\rm Re}\,a > {\rm Im}\,a$. 

Quite generally, experimental searches for $\eta$ nuclear quasibound states in 
proton, pion or photon induced $\eta$ production reactions are inconclusive. 
Regarding the onset of $\eta$ nuclear binding, Krusche and Wilkin~\cite{KW15} 
state: ``The most straightforward (but not unique) interpretation of the data 
is that $\eta d$ is unbound, $\eta\,^4$He is bound, but that the $\eta\,^3$He 
case is ambiguous." Indeed, with $\eta\,^3$He almost bound, one might expect 
that the denser $^4$He nucleus should help forming a bound $\eta\,^4$He. 
Nevertheless, a recent Faddeev-Yakubovsky evaluation \cite{FK17} of the 
scattering lengths $a_{\eta\,^A{\rm He}}$ for both He isotopes, $A=3,4$, 
finds this not to be the case, with the denser $^4$He apparently leading 
to a stronger reduction of the subthreshold $\eta N$ scattering amplitude 
than in $^3$He. 

The present overview reports and discusses recent few-body stochastic 
variational method (SVM) calculations of $\eta NNN$ and $\eta NNNN$ using 
several semi-realistic $NN$ interaction models together with two $\eta N$ 
interaction models that, perhaps, provide sufficient attraction to bind 
$\eta$ in the $^3$He and $^4$He isotopes~\cite{BFG15,BBFG17,BFG17}.

\section{$\eta N$ and $NN$ interaction model input}  
\label{sec:input} 

\begin{figure}[htb]
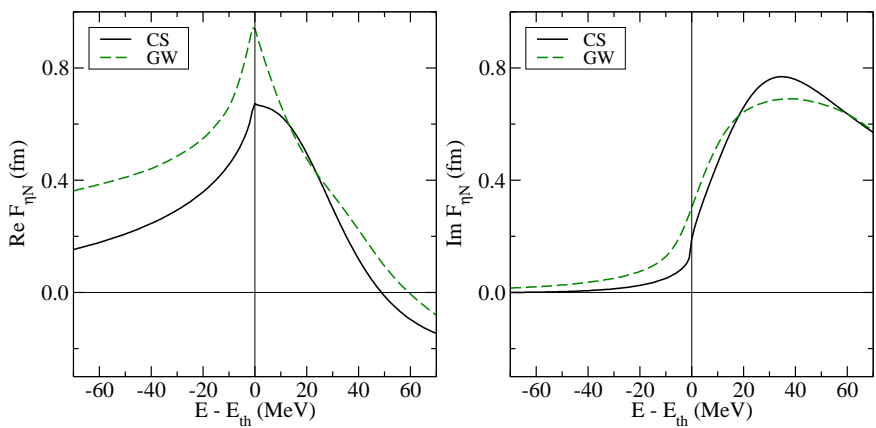
 
\centering 
\includegraphics[width=0.4\textwidth]{retamod2.eps} 
\includegraphics[width=0.4\textwidth]{ietamod2.eps} 
\caption{Real and imaginary parts of the $\eta N$ cm scattering amplitude 
near threshold in two meson-baryon coupled channel models: 
GW \cite{GW05} and CS \cite{CS13}.} 
\label{fig:GW&CS} 
\end{figure} 

\begin{figure}[htb]
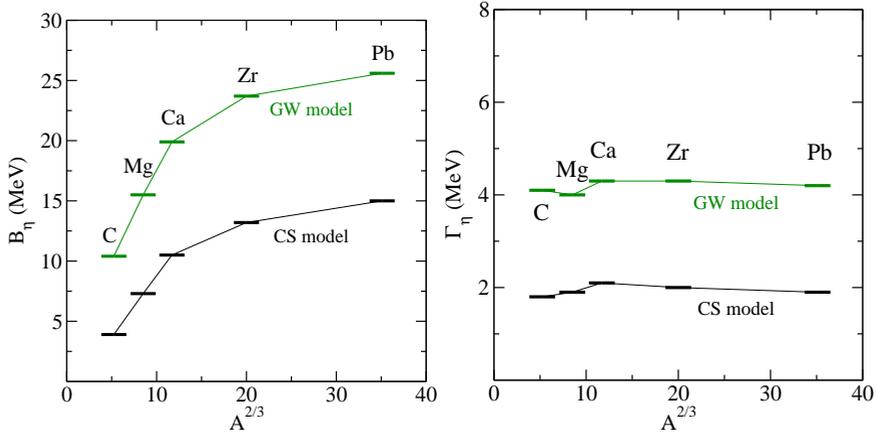
 
\centering 
\includegraphics[width=0.4\textwidth]{beta-all.eps} 
\includegraphics[width=0.4\textwidth]{gamma-all.eps} 
\caption{Binding energies $B_{\eta}$ (left) and widths $\Gamma_{\eta}$ 
(right) of $1s_{\eta}$ quasibound states across the periodic table calculated 
self consistently \cite{FGM13,CFGM14} using the GW and CS $\eta N$ scattering 
amplitudes of Fig.~\ref{fig:GW&CS}.} 
\label{fig:1s} 
\end{figure}

Figure \ref{fig:GW&CS} shows $\eta N$ $s$-wave scattering amplitudes 
$F_{\eta N}(E)$ calculated in two meson-baryon coupled-channel models across 
the $\eta N$ threshold where Re$\,F_{\eta N}$ has a cusp. These amplitudes 
exhibit a resonance about 50 MeV above threshold, the $N^{\ast}(1535)$. 
The sign of Re$\,F_{\eta N}$ below the resonance indicates attraction which 
is far too weak to bind the $\eta N$ two-body system. The threshold values 
$F_{\eta N}(E_{\rm th})$ are given by the scattering lengths 
\begin{equation} 
a_{\eta N}^{\rm GW}=(0.96+i0.26)~{\rm fm},~~~ 
a_{\eta N}^{\rm CS}=(0.67+i0.20)~{\rm fm},~~~ 
%a_{\eta N}^{\rm GR}=(0.26+i0.24)~{\rm fm}, 
\label{eq:GW&CS} 
\end{equation} 
with lower values below threshold ($E_{\rm th}=1487$~MeV). These free-space 
energy dependent subthreshold amplitudes are transformed to in-medium density 
dependent amplitudes, in terms of which optical potentials $V_{\eta}^{\rm opt}
(\rho)$ are constructed and used to calculate self consistently $\eta$ nuclear 
quasibound states. This procedure was applied in Refs.~\cite{FGM13,CFGM14} to 
several $\eta N$ amplitude models, with results for $1s_{\eta}$ quasibound 
states in models GW and CS shown in Fig.~\ref{fig:1s} from $^{12}$C to 
$^{208}$Pb. 

Figure~\ref{fig:1s} demonstrates that in both of these $\eta N$ amplitude 
models the $1s_{\eta}$ binding energy increases with $A$, saturating in 
heavy nuclei. Model GW, with larger $\eta N$ real and imaginary subthreshold 
amplitudes than in model CS, gives correspondingly larger values of $B_{\eta}$ 
and $\Gamma_{\eta}$. While model GW binds $\eta$ also in nuclei lighter than 
$^{12}$C (not shown in the figure) this needs to be confirmed in few-body 
calculations. 

\begin{figure}[htb] 
\centering 
\includegraphics[width=0.45\textwidth]{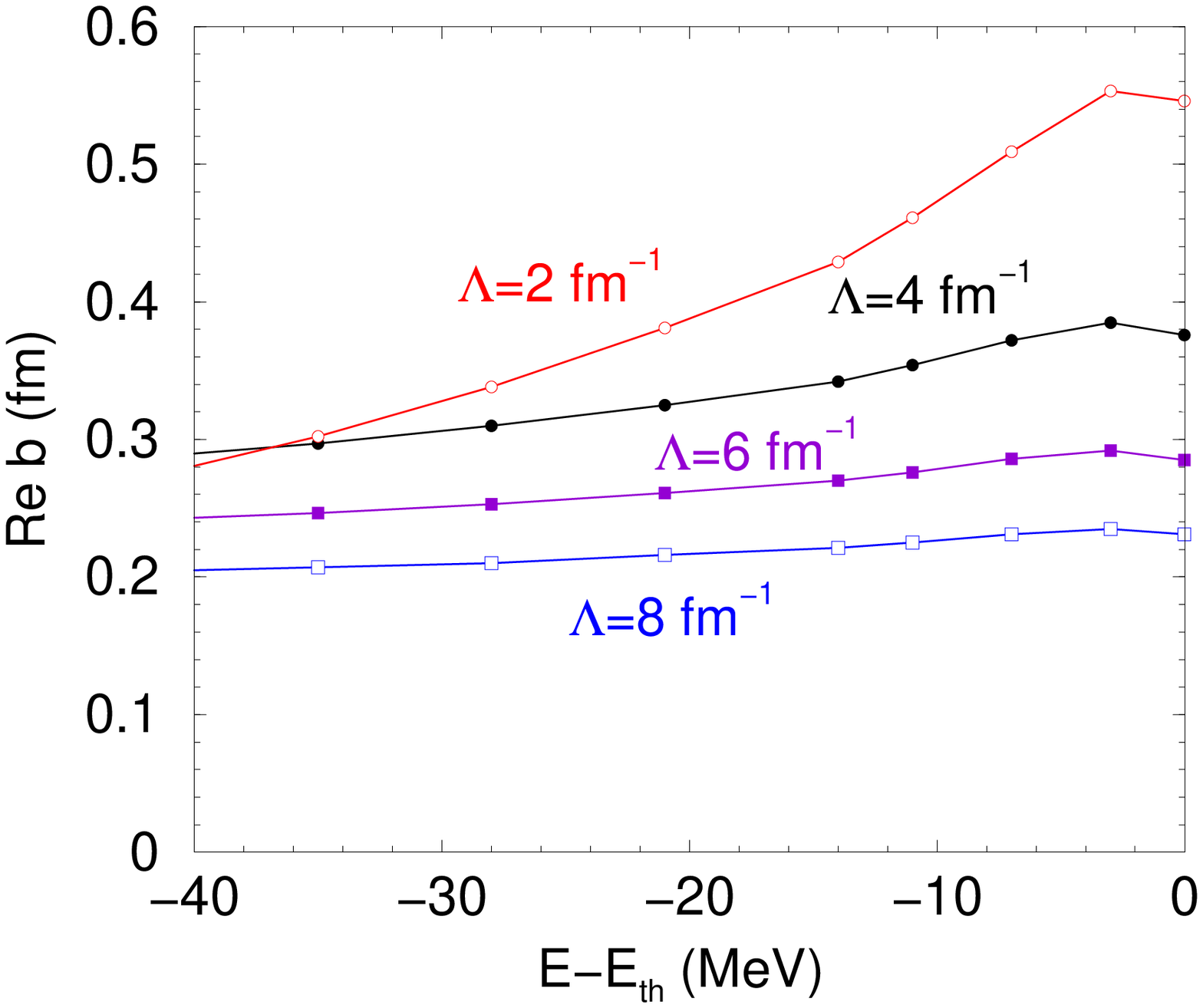} 
\includegraphics[width=0.45\textwidth]{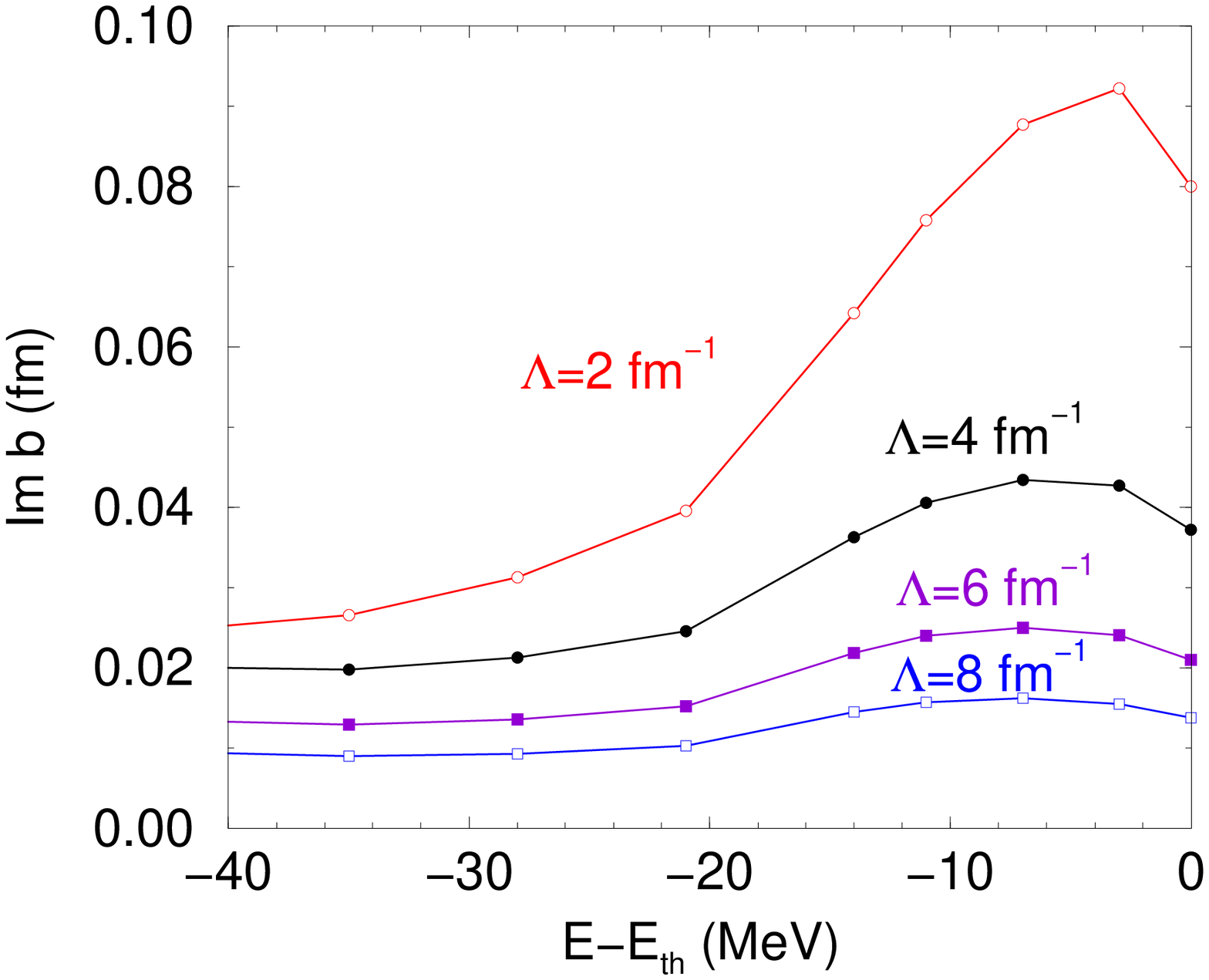} 
\caption{Real and imaginary parts of the strength function $b_{\Lambda}(E)$ 
of the effective $\eta N$ potential $v_{\eta N}^{\rm GW}(E)$, 
Eq.~(\ref{eq:v}), obtained from the scattering amplitude 
$F_{\eta N}^{\rm GW}(E)$ of Fig.~\ref{fig:GW&CS} below threshold 
for four values of the scale $\Lambda$~\cite{BBFG17}.} 
\label{fig:b(E)GW} 
\end{figure} 

Few-body calculations, in distinction from optical model calculations, require 
the use of effective $\eta N$ potentials $v_{\eta N}$ which reproduce the 
free-space $\eta N$ amplitudes below threshold. Fig.~\ref{fig:b(E)GW} shows 
subthreshold values of the energy dependent strength function $b_{\Lambda}(E)$ 
for $v_{\eta N}$ of the form 
\begin{equation} 
v_{\eta N}(E;r)=-\frac{4\pi}{2\mu_{\eta N}}b_{\Lambda}(E)\delta_{\Lambda}(r),
\,\,\,\,\,\,\,\delta_{\Lambda}(r)=\left(\frac{\Lambda}{2\sqrt{\pi}}
\right)^3\exp\left(-\frac{\Lambda^2 r^2}{4}\right), 
\label{eq:v} 
\end{equation} 
derived from the scattering amplitude $F_{\eta N}^{\rm GW}(E)$ of 
Fig.~\ref{fig:GW&CS} for several choices of inverse range $\Lambda$. 
The normalized Gaussian function $\delta_{\Lambda}(r)$ is perceived 
in \nopieft (pionless EFT) as a single $\eta N$ zero-range Dirac $\delta^{(3)}
({\bf r})$ contact term (CT), regulated by using a momentum-space scale 
parameter $\Lambda$. Regarding the choice of $\Lambda$, substituting 
the underlying short range vector-meson exchange dynamics by a single 
regulated CT suggests that the scale $\Lambda$ is limited to values 
$\Lambda\lesssim m_{\rho}$ ($\sim$4~fm$^{-1}$). 

Similarly, a \nopieft energy independent $v_{NN}(r)$ is derived at leading 
order (LO) by fitting a single regulated CT $\sim\delta_{\Lambda}(r)$ in each 
spin-isospin $s$-wave channel to the respective $NN$ scattering length. A $pp$ 
Coulomb interaction is included. To avoid $NNN$ and $\eta NN$ Thomas collapse 
in the limit $\Lambda\to\infty$, one introduces a three-body regulated CT for 
each of these three-body systems~\cite{BBFG17}: 
\begin{equation} 
V_{NNN}(r_{ij},r_{jk})=d_{NNN}^\Lambda\,\delta_{\Lambda}(r_{ij},r_{jk}),
\,\,\,\,\,\,\,
V_{\eta NN}(r_{i\eta },r_{\eta j})=d_{\eta NN}^\Lambda\,
\delta_{\Lambda}(r_{i\eta },r_{\eta j}), 
\label{eq:d} 
\end{equation} 
where $\delta_{\Lambda}(r_{ij},r_{jk}) = \delta_{\Lambda}(r_{ij})
\delta_{\Lambda}(r_{jk})$. 
The three-nucleon CT $d_{NNN}^\Lambda$ is fitted to $B_{\rm exp}(^3$He). 
With no further contact terms, $B_{\rm calc}(^4$He) is found in this \nopieft 
version~\cite{KPDB17} to vary moderately with $\Lambda$ and to exhibit 
renormalization scale invariance by approaching a finite value 
$B_{\Lambda\to\infty}(^4$He)=27.8$\pm$0.2~MeV that compares well with 
$B_{\rm exp}(^4$He)=28.3~MeV. In contrast, no $\eta$-related experimental 
datum is available for the $\eta NN$ CT $d_{\eta NN}^\Lambda$ to be fitted 
to. Two versions for choosing this CT were tested: (i) $d_{\eta NN}^\Lambda = 
d_{NNN}^\Lambda$, and (ii) setting $d_{\eta NN}^\Lambda$ so that $\eta d$ is 
just bound, i.e. $B_{\eta}(\eta d)=0$. Added to $v_{\eta N}^{\rm GW}(E)$, 
one finds that each of these versions prevents a potential collapse of 
$\eta d$, with calculated values of $B(\eta\,^A$He) that for $\Lambda\geq 
4$~fm$^{-1}$ are nearly independent of the adopted version, as shown in 
Fig.~\ref{fig:EFT} below.

\section{Energy independent \nopieft $\eta$ nuclear few-body calculations} 
\label{sec:thresh} 

\begin{figure}[htb] 
\centering 
\includegraphics[width=0.48\textwidth]{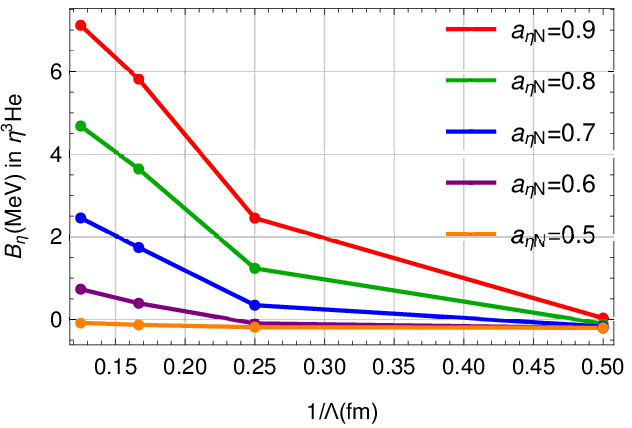} 
\includegraphics[width=0.48\textwidth]{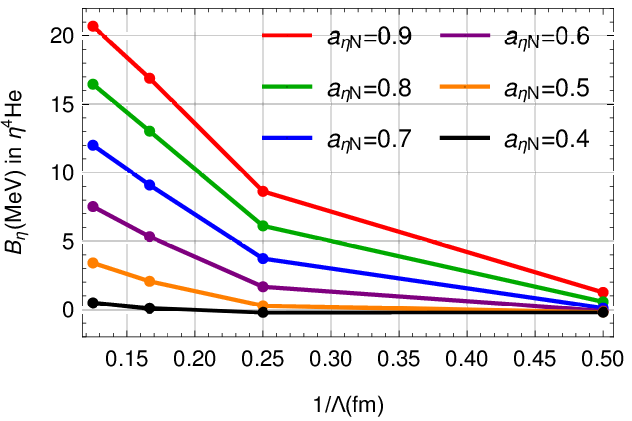} 
\caption{Separation energies $B_{\eta}$ obtained in SVM calculations of 
$\eta\,^3$He (left) and $\eta\,^4$He (right) using \nopieft $NN$ and $\eta N$ 
real interactions (\ref{eq:v}) fitted to values of $a_{\eta N}< 1$~fm, plus a 
universal $NNN$ and $\eta NN$ three-body CT (\ref{eq:d}), $d_{\eta NN}^\Lambda
=d_{NNN}^\Lambda$, as a function of $1/\Lambda$.} 
\label{fig:th} 
\end{figure} 

Fig.~\ref{fig:th} shows $\eta$ separation energies $B_{\eta}$ from \nopieft 
SVM calculations of $\eta\,^3$He and $\eta\,^4$He using energy independent 
$\eta N$ potentials $v_{\eta N}(E$=$E_{\rm th};r)$ fitted to a given real 
values of $a_{\eta N}$ for a few values of $\Lambda$. The figure suggests that 
binding $\eta\,^3$He ($\eta\,^4$He) requires that $a_{\eta N}\geq 0.55$~fm 
(0.45~fm), compatible with an effective value Re$\,a^{\prime}_{\eta N}$=0.48$
\pm$0.05~fm derived for a nearly bound $\eta\,^3$He \cite{Xie17}. For input 
values of $a_{\eta N}$ higher than shown in the figure, beginning at 
$a_{\eta N}$$\approx$1.2~fm, the calculated binding energies $B^{A=3,4}_{\eta}
(\Lambda >4$~fm$^{-1}$) diverge, apparently since $\eta d$ becomes bound then 
at $\Lambda$=4~fm$^{-1}$~\cite{BFG15}. Qualitative arguments in support of 
this $\eta d$ onset-of-binding value of $a_{\eta N}$ are given here in 
Appendix A.

\section{Energy dependence in $\eta$ nuclear few-body systems} 
\label{sec:sc} 

\begin{figure}[htb] 
\centering 
\includegraphics[width=0.5\textwidth]{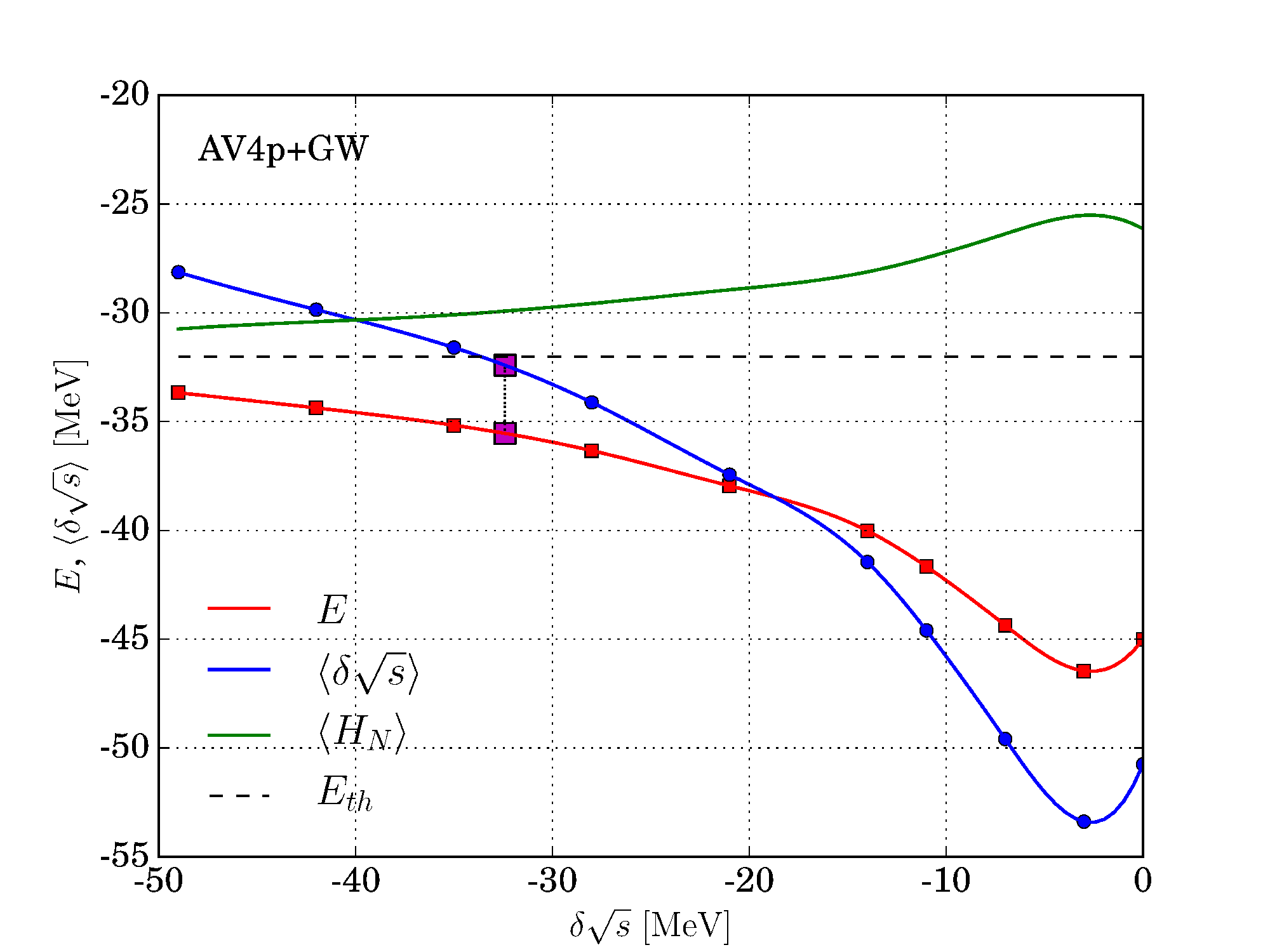} 
\includegraphics[width=0.4\textwidth, height=5cm]{srhocs.eps}
\caption{Left: $\eta\,^4$He bound state energy $E$ (red, squares) and the 
expectation value $\langle \delta \sqrt{s} \rangle$ (blue, circles), 
calculated using the AV4' $NN$ potential (denoted here AV4p), as a function 
of the input energy argument $\delta\sqrt{s}$ of the GW $\eta N$ potential 
with $\Lambda=4$~fm$^{-1}$. The dotted vertical line marks the self consistent 
output values of $\langle\delta\sqrt{s}\rangle$ and $E$. The horizontal dashed 
line denotes the calculated $^4$He g.s. energy, marking the threshold of 
$\eta$ binding. The green curve shows the expectation value $<H_N>$ of the 
nuclear core energy. Right: subthreshold $\eta N$ energies $\delta\sqrt{s}=
E-E_{\rm th}$ probed by the $\eta$ nuclear optical potential as a function of 
the relative nuclear RMF density in Ca. Each of the two curves was calculated 
self consistently for a particular $\eta N$ subthreshold amplitude model.} 
\label{fig:sc} 
\end{figure}

Having derived energy dependent $\eta N$ potentials $v_{\eta N}(E;r)$, see 
Eq.~(\ref{eq:v}) and Fig.~\ref{fig:b(E)GW}, a two-body subthreshold input 
energy $\delta\sqrt{s}\equiv E-E_{\rm th}$ needs to be chosen. However, 
$\delta\sqrt{s}$ is not conserved in the $\eta$ nuclear few-body problem, 
so the best one can do is to require that this choice agrees with the 
expectation value $\langle\,\delta\sqrt{s}\,\rangle$ generated in solving 
the few-body problem, as given by~\cite{BFG17} 
\begin{equation} 
\langle \delta\sqrt{s} \rangle = -\frac{B}{A}-\xi_{N}\frac{1}{A}
\langle T_A \rangle +\frac{A-1}{A} {\cal E}_{\eta} -\xi_A \xi_{\eta}
\left ( \frac{A-1}{A}\right )^2\langle T_{\eta} \rangle. 
\label{eq:deltas} 
\end{equation} 
Here $\xi_{N(\eta)}=m_{N(\eta)}/(m_N+m_{\eta}),~\xi_A=Am_N/(Am_N+m_{\eta})$, 
$T_A$ and $T_{\eta}$ denote the nuclear and $\eta$ kinetic energy operators 
in appropriate Jacobi coordinates, $B$ is the total binding energy, and 
${\cal E}_{\eta}=\langle H-H_N \rangle$ with each Hamiltonian defined in its 
own cm frame. Self consistency (SC), $\langle\,\delta\sqrt{s}\,\rangle = 
\delta\sqrt{s}$, is imposed in our calculations, as demonstrated graphically 
in Fig.~\ref{fig:sc} (left). Applications of SC to meson-nuclear systems are 
reviewed in Ref.~\cite{Gal14}. For recent $K^-$-atom and nuclear applications 
see Refs.~\cite{FG17,HM17}. More recently, Hoshino et al.~\cite{Jap17} argued 
in a $K^-d$ study that by applying this procedure one violates the requirement 
of total momentum conservation. In Appendix B here we show specifically for 
$A=2$ that our choice of SC Eq.~(\ref{eq:deltas}) is not in conflict with any 
conservation law. 

Finally, we note that Eq.~(\ref{eq:deltas}) in the limit $A >> 1$ coincides 
with the optical model downward energy shift (supplemented by a Coulomb term) 
used in recent $K^-$ atom and nuclear studies~\cite{FG17,HM17}: 
\begin{equation} 
\langle \delta\sqrt{s}\rangle = -B_N\,\frac{\rho}{\bar\rho}-
\xi_N B_{\eta}\,\frac{\rho}{\rho_0}-\xi_N T_N\,(\frac{\rho}{\bar\rho})^{2/3}+
\xi_{\eta}{\rm Re}\,V_{\eta}^{\rm opt}(\delta\sqrt{s}), 
\label{eq:opt} 
\end{equation} 
where $T_N=\langle T_A \rangle/A=23.0$~MeV at the average nuclear density 
$\bar\rho$, $B_N=B_{\rm nuc}/A\approx 8.5$~MeV is an average nucleon binding 
energy and $B_{\eta}$ denotes the calculated $\eta$ separation energy. All 
terms here are negative, thereby leading to a downward energy shift. 

The SC procedure is demonstrated in Fig.~\ref{fig:sc} (left) for $\eta\,^4$He 
binding energy calculated using the AV4' $NN$ potential and GW $\eta N$ 
potential with $\Lambda$=4~fm$^{-1}$. The $\eta\,^4$He bound state energy $E$ 
(excluding rest masses) 
and the output expectation value $\langle\delta\sqrt{s}\rangle$, where $\delta
\sqrt{s}$ stands for the $\eta N$ cm energy with respect to its threshold 
value $E_{\rm th}$, are plotted as a function of the subthreshold input energy 
argument $\delta\sqrt{s}$ of the potential $v^{\rm GW}_{\eta N}$. The SC 
condition requires $\delta\sqrt{s}=\langle\delta\sqrt{s}\rangle$ which is 
satisfied at $-$32.4~MeV. The corresponding value of $E(\langle\delta\sqrt{s}
\rangle)$ then represents the SC energy of $\eta\,^4$He, with $B^{\rm SC}_{
\eta}=3.5$~MeV, considerably less than the value $B^{\rm th}_{\eta}=13$~MeV 
obtained by disregarding the energy dependence of $v^{\rm GW}_{\eta N}$ and 
using its threshold value corresponding to $\delta\sqrt{s}=0$. 

In Fig.~\ref{fig:sc} (right) we present the $\eta N$ downward energy shift 
$\delta\sqrt{s}=E-E_{\rm th}$ as a function of the relative nuclear density 
$\rho/\rho_0$ in Ca, evaluated self consistently via Eq.~(\ref{eq:opt}) in the 
CS and GW models. The energy shift at $\rho_0$ is $-55\pm 10$~MeV, about twice 
larger than the SC condition $\delta\sqrt{s}=-B_{\eta}$ applied in some other 
works, e.g.~\cite{GR02}. The GW shift exceeds the CS shift owing to the 
stronger GW amplitude of Fig.~\ref{fig:GW&CS} and both were incorporated in 
the calculation of $1s_{\eta}$ quasibound nuclear states, Fig.~\ref{fig:1s}.

\section{Results of $\eta$ nuclear few-body calculations} 
\label{sec:res} 

\begin{figure}[htb] 
\centering 
\includegraphics[width=0.48\textwidth]{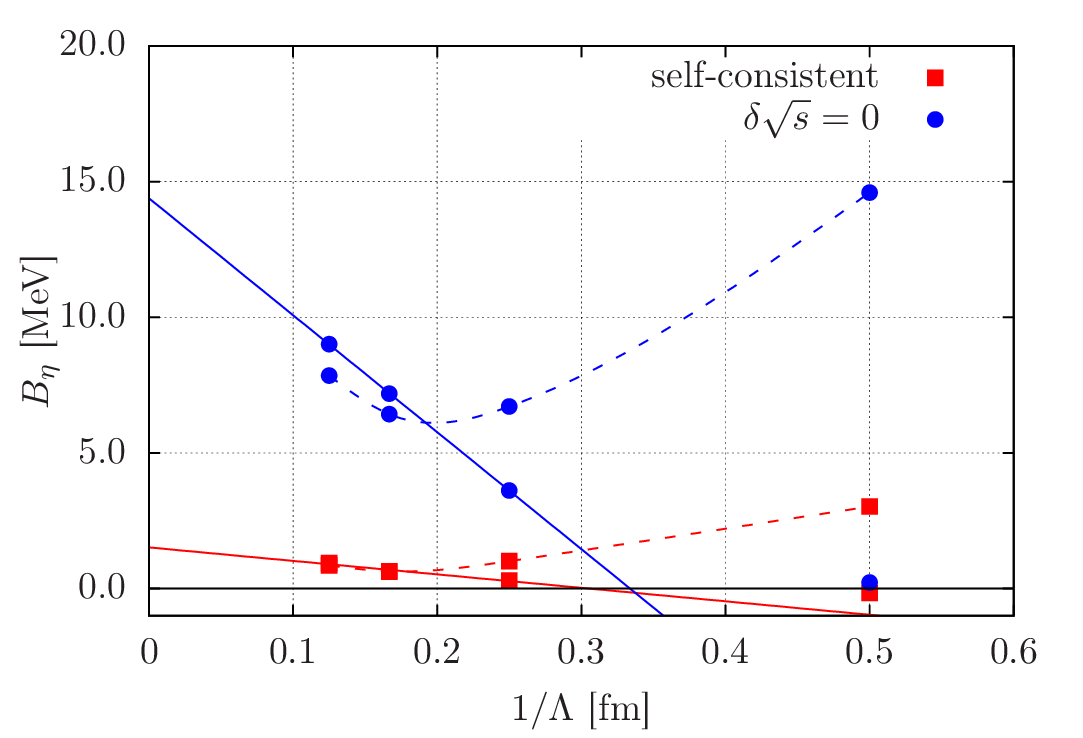} 
\includegraphics[width=0.48\textwidth]{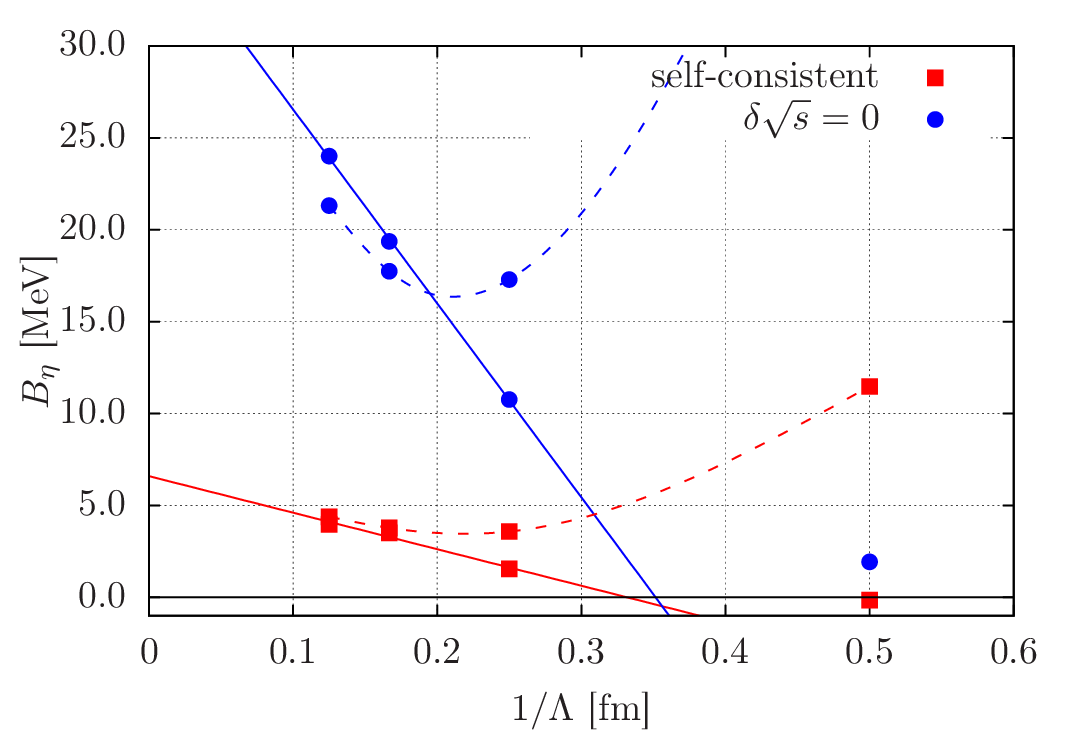} 
\caption{$B_{\eta}(\eta\,^3$He) (left) and $B_{\eta}(\eta\,^4$He) (right) as 
a function of $1/\Lambda$ from \nopieft few-body calculations~\cite{BBFG17} 
using $v_{\eta N}^{\rm GW}$, with (squares) \& without (circles) imposing 
self consistency. Solid lines: $d_{\eta NN}^\Lambda = d_{NNN}^\Lambda$, 
dashed lines: $d_{\eta NN}^\Lambda$ ensuring that $B_{\eta}(\eta d)=0$.} 
\label{fig:EFT} 
\end{figure}

\begin{figure}[htb] 
\centering 
\includegraphics[width=0.48\textwidth]{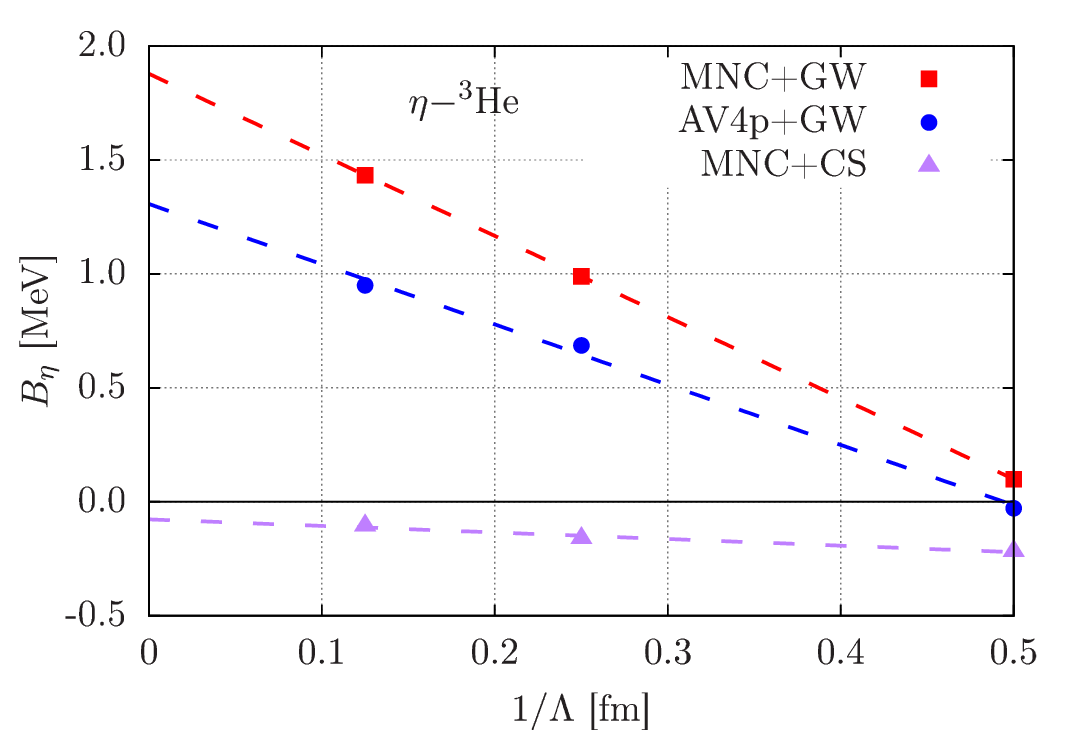} 
\includegraphics[width=0.48\textwidth]{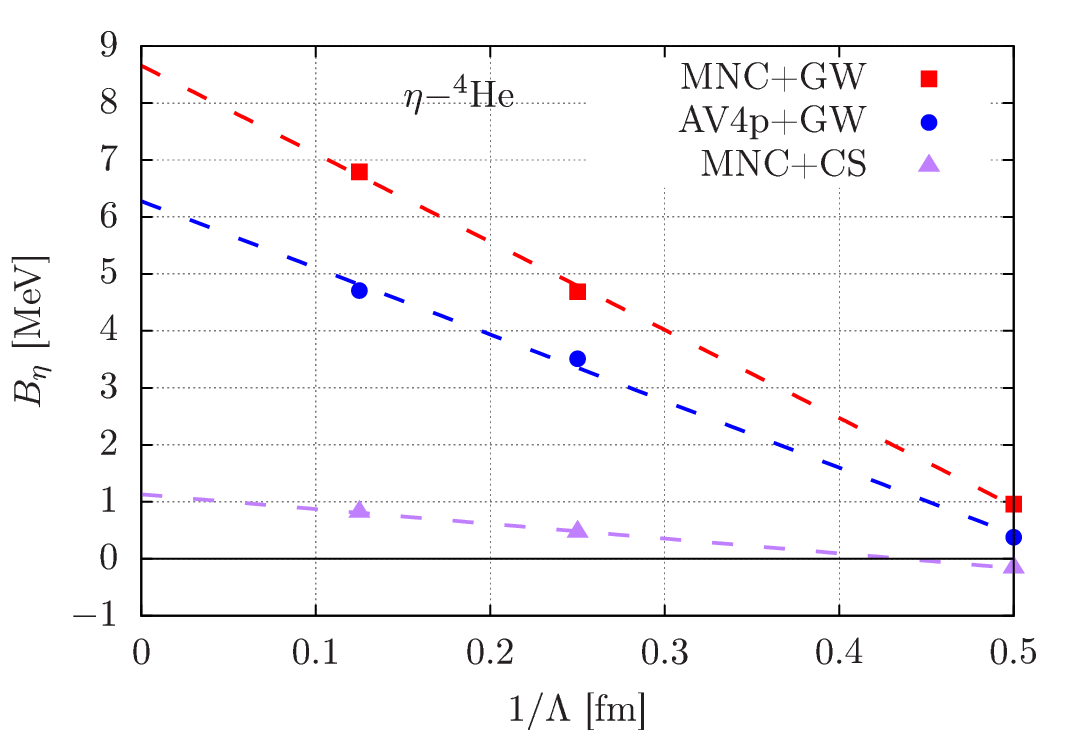} 
\caption{$B_{\eta}(\eta\,^3$He) (left) and $B_{\eta}(\eta\,^4$He) (right) as 
a function of $1/\Lambda$ from few-body calculations~\cite{BFG17} using $NN$ 
and $\eta N$ interactions, as marked, and imposing self consistency.} 
\label{fig:AV4MNC}  
\end{figure}

Our fully self consistent $\eta NN$, $\eta NNN$ and $\eta NNNN$ bound-state 
calculations~\cite{BFG15,BBFG17,BFG17} use the following nuclear core models: 
(i) \nopieft including a three-body contact term~\cite{KPDB17}, (ii) AV4p, 
a Gaussian basis adaptation of the Argonne AV4' $NN$ potential~\cite{AV4}, 
and (iii) MNC, the Minnesota soft core $NN$ potential~\cite{MNC}. Models 
GW~\cite{GW05} and CS~\cite{CS13} were used to generate energy dependent 
$\eta N$ potentials which prove too weak to bind any $\eta NN$ system when 
using AV4p or MNC for the nuclear core model. Calculated $\eta$ separation 
energies $B_{\eta}$ are shown in Figs.~\ref{fig:EFT} and~\ref{fig:AV4MNC}. 

Fig.~\ref{fig:EFT} demonstrates in \nopieft the moderating effect that 
imposing SC (red, squares) by using $v_{\eta N}^{\rm GW}(E_{\rm sc})$, rather 
than using threshold values $v_{\eta N}^{\rm GW}(E_{\rm th})$ (blue, circles), 
bears on the calculated $B_{\eta}$ values and their $\Lambda$ scale 
dependence~\cite{BBFG17}. Near $\Lambda$=4~fm$^{-1}$, imposing sc lowers 
$B_{\eta}(\eta\,^3$He) by close to 5 MeV and $B_{\eta}(\eta\,^4$He) by close 
to 10 MeV. The figure demonstrates that $B_{\eta}(\eta\,^4$He) is always 
larger than $B_{\eta}(\eta\,^3$He). Focusing on scale parameters near 
$\Lambda$=4~fm$^{-1}$ one observes that $\eta\,^3$He is hardly bound by 
a fraction of MeV, whereas $\eta\,^4$He is bound by a few MeV. The choice 
of three-body CT $d_{\eta NN}^\Lambda$ hardly matters for $\Lambda > 4
$~fm$^{-1}$, becoming substantial at $\Lambda < 4$~fm$^{-1}$. 

Fig.~\ref{fig:AV4MNC} demonstrates in non-EFT calculations the dependence of 
$B_{\eta}$, calculated self consistently, on the choice of $NN$ and $\eta N$ 
interaction models. Using the more realistic AV4' $NN$ interaction results 
in less $\eta$ binding than using the soft-core MNC $NN$ interaction. For 
$v_{\eta N}^{\rm GW}$ near $\Lambda$=4~fm$^{-1}$ the difference amounts to 
about 0.3~MeV for $\eta\,^3$He and about 1.5~MeV for $\eta\,^4$He; $\eta\,^3
$He appears then barely bound whereas $\eta\,^4$He is bound by a few MeV. 
The weaker $v_{\eta N}^{\rm CS}$ does not bind $\eta\,^3$He and barely binds 
$\eta\,^4$He using the MNC $NN$ interaction, implying that $\eta\,^4$He is 
unlikely to bind for the more realistic AV4' $NN$ interaction. For smaller, 
but still physically acceptable values of $\Lambda$ down to $\Lambda = 2
$~fm$^{-1}$, $\eta\,^3$He becomes unbound and $\eta\,^4$He is barely bound 
using the AV4' $NN$ and GW $\eta N$ interactions. 

The $B_{\eta}$ values calculated in Refs.~\cite{BFG15,BBFG17,BFG17} were 
calculated assuming real Hamiltonians, justified by Im$\,v_{\eta N}$$\ll$Re$
\,v_{\eta N}$ from Fig.~\ref{fig:b(E)GW}. This approximation is estimated to 
add near threshold less than 0.3~MeV to $B_{\eta}$. Perturbatively-calculated 
widths $\Gamma_{\eta}$ of weakly bound states amount to only few MeV, 
outdating those reported in Ref.~\cite{BFG15}. 

\begin{figure}[htb] 
\centering 
\includegraphics[width=0.7\textwidth]{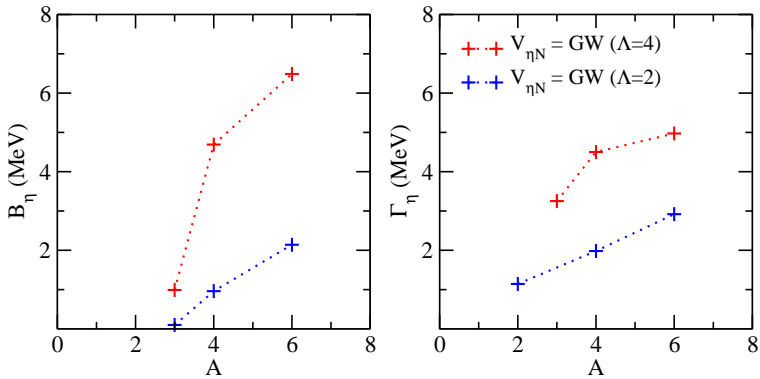} 
\caption{Preliminary SVM results for binding energies $B_{\eta}$ (left) and 
widths $\Gamma_{\eta}$ (right) of $1s_{\eta}$ quasibound states in $^3$He, 
$^4$He and $^6$Li, calcualted using the Minnesota $NN$ potential and the GW 
$\eta N$ potential for $\Lambda = 2$ and 4~fm$^{-1}$.} 
\label{fig:martin} 
\end{figure} 

In future work it will be interesting to extend the present SVM few-body 
calculations to heavier nuclei, beginning with light $p$-shell nuclei. 
This represents highly non-trivial task. In Fig.~\ref{fig:martin} we present 
preliminary results for $\eta\,^6$Li, using the central Minnesota $NN$ and GW 
$\eta N$ potentials. In this calculation the $^6$Li nuclear core consisted of 
a single $S=1,T=0$ spin-isospin configuration, yielding $B(^6$Li)=34.66~MeV 
which is short by almost 2~MeV with respect to a calculation reported in 
Ref.~\cite{NC14} that used the same $NN$ interaction while including more 
spin-isospin configurations. The figure suggests that $\eta\,^6$Li is 
comfortably bound, even for as low value of scale parameter as $\Lambda 
= 2$~fm$^{-1}$.

\section{Summary} 
\label{sec:sum} 

Based mostly on the AV4' results in Fig.~\ref{fig:AV4MNC}, which are close to 
the \nopieft results in Fig.~\ref{fig:EFT}, we conclude that $\eta\,^3$He 
becomes bound for Re$\,a_{\eta N}\sim 1$~fm, as in model GW, while $\eta\,^4
$He binding requires a lower value of Re$\,a_{\eta N}\sim 0.7$~fm, almost 
reached in model CS. These Re$\,a_{\eta N}$ onset values, obtained by 
incorporating the requirements of $\eta N$ subthreshold kinematics, are 
obviously \textit{larger} than those estimated in Sect.~\ref{sec:thresh} 
upon calculating with $v_{\eta N}(E=E_{\rm th};r)$ threshold input. Finally, 
Re$\,a_{\eta N} < 0.7$~fm if $\eta\,^4$He is unbound, as might be deduced 
from the recent WASA-at-COSY search~\cite{AAB16}.

\section*{Appendix A: Onset of $\eta d$ binding} 

Here we apply the Brueckner formula~\cite{Bru53}, expressing the $\eta d$ 
scattering length in terms of the $\eta N$ scattering length, to discuss 
qualitatively the onset of $\eta d$ binding. This formula was originally 
proposed for a system of a light meson ($\pi$ meson) and two heavy static 
nucleons. More recently it was used to estimate the $K^- d$ scattering length 
(see derivation and discussion in Ref.~\cite{Gal07}) where the meson-nucleon 
mass ratio is similar to that for $\eta N$. For $\eta d$ the Brueckner formula 
assumes the form 
\begin{equation} 
\label{eq:int} 
a_{\eta d}=\int{a_{\eta d}(r){|\psi_d({\bf r})|}^2 {\rm d}{\bf r}}~, 
\end{equation} 
\begin{equation} 
\label{eq:bru} 
a_{\eta d}(r) = {\left (1+\frac{m_{\eta}}{m_d} \right )}^{-1}\,
\frac{{\tilde a}_p + {\tilde a}_n + 2 {\tilde a}_p {\tilde a}_n /r} 
{1-{\tilde a}_p {\tilde a}_n /r^2}~, 
\end{equation} 
where $\tilde a = (1+m_{\eta}/m_N) a$, with $a_p$ and $a_n$ standing for 
$a_{\eta p}$ and $a_{\eta n}$ respectively in the $\eta N$ cm system. The 
numerator in the Brueckner formula consists of single- and double-scattering 
terms, whereas the denominator provides for the renormalization of these 
terms by higher-order scattering terms. Since $a_p=a_n$ for the isoscalar 
$\eta$ meson, Eq.~(\ref{eq:bru}) reduces to a simpler form, 
\begin{equation} 
\label{eq:deuteron} 
a_{\eta d}(r) = \frac{2}{1+\frac{m_{\eta}}{m_d}}\,\frac{{\tilde a}_{\eta N}} 
{1-{\tilde a}_{\eta N}/r}~,   
\end{equation} 
which leads to the following approximate expression: 
\begin{equation} 
\label{eq:etad}
a_{\eta d} = \frac{2}{1+\frac{m_{\eta}}{m_d}}\,\frac{{\tilde a}_{\eta N}}
{1-{\tilde a}_{\eta N}\langle 1/r \rangle_d}~, 
\end{equation} 
with expansion parameter ${\tilde a}\,\langle 1/r\rangle_d$, where $\langle 
1/r \rangle_d \approx 0.45~{\rm fm}^{-1}$ for a realistic deuteron 
wavefunction~\cite{VAr06}. Hence, this multiple scattering series faces 
divergence for sufficiently large $\eta N$ scattering length, say $a > 1.4$~fm. 

Several straightforward applications of Eq.~(\ref{eq:etad}) are as follows: 
\begin{itemize} 
\item 
For Re$\,a_{\eta N}^{\rm GW}=0.96$~fm, suppressing Im$\,a_{\eta N}^{\rm GW}$, 
one gets $a_{\eta d}=7.46$~fm. Increasing this GW input value of $a_{\eta N}$, 
a critical value $a^{\rm crit}_{\eta N}=1.40$~fm is reached at which the 
denominator in Eq.~(\ref{eq:etad}) vanishes, signaling the appearance of 
a zero-energy $\eta d$ bound state. 
\item 
The LO \nopieft nuclear calculations~\cite{KPDB17} yield a more compact 
deuteron, $r_{\rm rms}=1.55$~fm for $\Lambda\to\infty$ compared to the 
'experimental' value $r_{\rm rms}=1.97$~fm. Scaling the value $\langle 
1/r \rangle_d = 0.45$ used in Eq.~(\ref{eq:etad}) by 1.97/1.55, 
one gets $a_{\eta d}=18.1$~fm and $a^{\rm crit}_{\eta N}=1.10$~fm. 
\item 
For the fully complex scattering length $a_{\eta N}^{\rm GW}=0.96+i0.26$~fm, 
one gets $a_{\eta d}=4.66+i4.76$~fm. Increasing Re$\,a_{\eta N}$ at a frozen 
value of Im$\,a_{\eta N}$, Re$\,a_{\eta d}$ reverses its sign at Re$\,a^{\rm 
crit}_{\eta N}=1.35$~fm while Im$\,a_{\eta d}$ keeps positive all through. 
\item 
At Re$\,a^{\rm crit}_{\eta N}=1.59$~fm, $|$Re$\,a_{\eta d}|$ becomes larger 
than Im$\,a_{\eta d}$, which signals a threshold $\eta d$ bound state. 
\end{itemize}

\section*{Appendix B: $\eta N$ subthreshold kinematics} 

Here we outline the choice of the $\eta N$ subthreshold energy shift 
$\delta\sqrt{s}\equiv\sqrt{s_{\eta N}}-(m_N+m_{\eta})$ applied in our $\eta$ 
nuclear few-body works~\cite{BFG15,BBFG17,BFG17}, see Eq.~(\ref{eq:deltas}), 
with emphasis on the three-body $\eta d$ system. Since the $\eta N$ effective 
potential $v_{\eta N}$ discussed in Sect.~\ref{sec:input} is energy dependent, 
one needs to determine as consistently as possible a fixed {\it input} value 
$\delta\sqrt{s}$ at which $v_{\eta N}$ should enter the $\eta$ nuclear 
few-body calculation. The two-body Mandelstam variable $\sqrt{s_{\eta N}}=
\sqrt{(E_{\eta}+E_N)^2-({\vec p}_{\eta}+{\vec p}_N)^2}$ which reduces to 
$(E_{\eta}+E_N)$ in the $\eta N$ two-body cm system is not a conserved 
quantity in the $\eta$ nuclear few-body problem since spectator nucleons 
move the interacting $\eta N$ two-body subsystem outside of its cm system. 
We proceed to evaluate the expectation value of {\it output} 
values of $\delta\sqrt{s}$, replacing $\sqrt{s_{\eta N}}$ by 
$(1/A){\sum_{i=1}^{A}\sqrt{(E_{\eta}+E_i)^2-({\vec p}_{\eta}+{\vec p}_i)^2}}$ 
due to the antisymmetry of the nuclear wavefunction.  
Expanding about the $\eta N$ threshold, one gets in leading order of $p^2$ 
\begin{equation} 
\langle\,\delta\sqrt{s}\,\rangle\approx\frac{1}{A}\,\langle\,
{\sum_{i=1}^{A}({\cal E}_{\eta}+{\cal E}_i)}
-{\sum_{i=1}^{A}\frac{({\vec p}_{\eta}+{\vec p}_i)^2}{2(m_N+m_{\eta})}}\,
\rangle, 
\label{eq:delsqrts} 
\end{equation} 
where ${\cal E}_{\eta}=E_{\eta}-m_{\eta}$ and ${\cal E}_i=E_i-m_N$. Since 
${\sum_{i=1}^{A}{\cal E}_i}$ is naturally identified with the expectation 
value of the nuclear Hamiltonian $H_N$, ${\sum_{i=1}^{A}{\cal E}_i}=\langle 
H_N \rangle = E_{\rm nuc} = -B_{\rm nuc}$, it is natural and also consistent 
to identify ${\cal E}_{\eta}$ with the expectation value of $(H-H_N)$, 
${\cal E}_{\eta}=\langle H-H_N \rangle$. Furthermore, recalling that 
${\cal E}_{\eta}+{\sum_{i=1}^{A}{\cal E}_i}$=$E$=$-B$, where $E=\langle H 
\rangle$ is the total $\eta$ nuclear energy and $B$ is the total binding 
energy, the sum over the momentum independent part in Eq.~(\ref{eq:delsqrts}) 
gives $[-B+(A-1){\cal E}_{\eta}]/A$, thereby reproducing two of the four terms 
in Eq.~(\ref{eq:deltas}). Note that ${\cal E}_{\eta}$ is negative and its 
magnitude exceeds the $\eta$ separation energy $B_{\eta}$. The sum over the 
momentum dependent part of Eq.~(\ref{eq:delsqrts}) yields the other two terms 
of Eq.~(\ref{eq:deltas}), which we demonstrate for $\eta d$, $A=2$. 

Since the $\eta d$ calculation employes translationally invariant coordinate 
sets, the total momentum vanishes sharply: 
$({\vec p}_{\eta}+{\vec p}_1+{\vec p}_2)=0$. We then substitute 
${{\vec p}_1}^{\,2}$ for $({\vec p}_{\eta}+{\vec p}_2)^{\,2}$ and 
${{\vec p}_2}^{\,2}$ for $({\vec p}_{\eta}+{\vec p}_1)^{\,2}$ in the momentum 
dependent part in Eq.~(\ref{eq:delsqrts}), resulting in momentum dependence 
proportional to ${{\vec p}_1}^{\,2}+{{\vec p}_2}^{\,2}$. This is rewritten as 
\begin{equation} 
{{\vec p}_1}^{\,2}+{{\vec p}_2}^{\,2}=\frac{1}{2}[({\vec p}_1-{\vec p}_2)^2+ 
({\vec p}_1+{\vec p}_2)^2]=2{{\vec p}_{N:N}}^{\,2}+\frac{1}{2}
{{\vec p}_{\eta}}^{\,2}, 
\label{eq:incoherent} 
\end{equation} 
where ${\vec p}_{N:N}$ is the nucleon-nucleon relative momentum operator. 
To obtain the $\eta$ momentum operator ${\vec p}_{\eta}$ on the r.h.s. 
we used again total momentum conservation. Finally, transforming 
${{\vec p}_{N:N}}^{\,2}$ and ${{\vec p}_{\eta}}^{\,2}$ to intrinsic 
kinetic energies, $T_{N:N}$ for the internal motion of the deuteron 
core and $T_{\eta}$ for that of the $\eta$ meson with respect to the 
NN cm, one gets for this $A=2$ special case  
\begin{equation} 
\langle\,\delta\sqrt{s}\,\rangle_{\eta d}\approx -\frac{1}{2}\left 
(B-{\cal E}_{\eta}+\xi_N\langle T_{N:N}\rangle+\xi_{A=2}\xi_{\eta}\frac{1}{2}
\langle T_{\eta} \rangle \right ), 
\label{eq:delsetad} 
\end{equation} 
which agrees with Eq.~(\ref{eq:deltas}) for $A=2$ upon realizing that 
$T_{N:N}$ here coincides with $T_{A=2}$ there. To get idea of the relative 
importance of the various terms in this expression, we assume a near-threshold 
$\eta d$ bound state for which both ${\cal E}_{\eta}$ and $\langle T_{\eta} 
\rangle$ are negligible (fraction of MeV each) and $B\to B_d\approx 2.2$~MeV. 
With $\langle T_{N:N} \rangle \to \langle T_d \rangle$, and with a deuteron 
kinetic energy $\langle T_d \rangle$ in the range of 10 to 20 MeV, this term 
provides the largest contribution to the downward energy shift which is then 
of order $-$5~MeV for the diffuse deuteron nuclear core.

\section*{Acknowledgments} 

The work of A.C., J.M. and M.S. was supported by the GACR Grant 
No. P203/15/04301S.

\end{document}